\documentclass[%
 aip,
 pof,
 amsmath,amssymb,
preprint,%
]{revtex4-2}

\usepackage{graphicx}
\usepackage{dcolumn}
\usepackage{bm}
\usepackage[mathlines]{lineno}
\linenumbers\relax 
\usepackage{wasysym}
\usepackage[utf8]{inputenc}
\usepackage[T1]{fontenc}
\usepackage{etoolbox}
\usepackage{siunitx}
\usepackage{xcolor}
\usepackage{pifont}
\graphicspath{{./pic/}}

\usepackage{siunitx}
\usepackage[final]{changes}
\usepackage{hyperref}
\usepackage{csquotes}

\let\oldcitep\citep
\renewcommand{\citep}[1]{\mbox{\oldcitep{#1}}}
\let\oldcitet\citet
\renewcommand{\citet}[1]{\mbox{\oldcitet{#1}}}
\let\oldcite\cite
\renewcommand{\cite}[1]{\mbox{\oldcite{#1}}}

\usepackage{CJK}
\newcommand{\upd}{\si{\,d}}

\makeatletter
\def\@email#1#2{%
 \endgroup
 \patchcmd{\titleblock@produce}
  {\frontmatter@RRAPformat}
  {\frontmatter@RRAPformat{\produce@RRAP{*#1\href{mailto:#2}{#2}}}\frontmatter@RRAPformat}
  {}{}
}%
\makeatother

\begin{document}

\nolinenumbers
\begin{CJK*}{UTF8}{gbsn}

\title{Hidden Turbulence in van Gogh's \textbf{\textit{The Starry Night}}}

\author{Yinxiang Ma (马寅翔)}
\affiliation{State Key Laboratory of Marine Environmental Science \& College of Ocean and Earth Sciences,
Xiamen University, Xiamen, China}


\author{Wanting Cheng (程婉婷)}
\affiliation{Center for Complex Flows and Soft Matter Research and Department of Mechanics and Aerospace Engineering, Southern University of Science and Technology, Shenzhen, Guangdong, China}

\author{Shidi Huang (黄仕迪)}
\affiliation{Center for Complex Flows and Soft Matter Research and Department of Mechanics and Aerospace Engineering, Southern University of Science and Technology, Shenzhen, Guangdong, China}

\author{Fran\c{c}ois G. Schmitt}
\affiliation{CNRS, Univ. Lille, Univ. Littoral Cote d'Opale, UMR 8187, LOG, Laboratoire d'Oc\'eanologie et de G\'eosciences, F 62930 Wimereux, France}

\author{Xin Lin (林昕)}
\affiliation{State Key Laboratory of Marine Environmental Science \& College of Ocean and Earth Sciences,
Xiamen University, Xiamen, China}

\author{Yongxiang Huang (黄永祥)}
\email{yongxianghuang@\{gmail.com,xmu.edu.cn\}}
\affiliation{State Key Laboratory of Marine Environmental Science \& College of Ocean and Earth Sciences,
Xiamen University, Xiamen, China}
\affiliation{Fujian Engineering Research Center for Ocean Remote Sensing Big Data, Xiamen, China}
\affiliation{Center for Marine Meteorology and Climate Change, Xiamen University, Xiamen China}

\date{\today}

\begin{abstract}

Turbulent skies have often inspired artists, particularly in the iconic swirls of Vincent van Gogh's \textbf{\textit{The Starry Night}}. For an extended period, debate has raged over whether the flow pattern in this masterpiece adheres to Kolmogorov's theory of turbulence. In contrast to previous studies that examined only part of this painting, {\textit{all and only the}} whirls/eddies in the painting are taken into account in this work, following the Richardson-Kolmogorov's cascade picture of turbulence. Consequently, the luminance's Fourier power spectrum spontaneously exhibits a characteristic $-5/3$ Kolmogorov-like power-law. This result suggests that van Gogh had a very careful observation of real flows, so that not only the sizes of whirls/eddies in \textbf{\textit{The Starry Night}} but also their relative distances and intensity follow the physical law that governs turbulent flows.
Moreover, a "$-1$"-like power-law persists in the spectrum below the scales of the smallest whirls, hinting at Batchelor-type scalar turbulence with a high Schmidt number. Our study thus unveils the hidden turbulence captured within \textbf{\textit{The Starry Night}}.


\end{abstract}

\maketitle
\end{CJK*}

\section{Introduction}



Turbulent flows or flow patterns similar to turbulence are ubiquitous in nature, ranging from atmospheric and oceanic flows of planetary-scale\citep{Frisch1995} to high-concentration bacteria suspensions at micro-scales.\citep{Wensink2012PNAS,Qiu2016PRE} One common feature of these phenomena is the existence of abundant swirling structures, which are also well captured by many artists and become key elements in their paintings. Examples include \textbf{\textit{The Yellow River Breaches Its Course}} attributed to 13th-century Chinese artist Yuan Ma,\citep{Zhou2021PR,Warhaft2022AS} a series of drawings of water flows by Leonardo da Vinci in 1500s,\citep{Frisch1995,Chen2019ME,Raissi2020Science,Marusic2021ARFM,Colagrossi2021PoF} \textbf{\textit{The Great Wave off Kanagawa}} by Katsushika Hokusai in 1831,\citep{Cartwright2009NRRS,Dudley2013NRRS,Ornes2014PNAS} 
and \textbf{\textit{The Starry Night}} by Vincent van Gogh in 1890,\citep{Aragon2008JMIV,Olson2014book,Beattie2019arXiv,Finlay2020JoT,Spreafico2021JMA,Sherman2023PP} to name a few. Turbulence-like patterns appearing in these artworks have inspired scientists to examine how close these patterns are to real turbulent flows. In this regard, an interesting but unsettled debate is whether the swirling structures in van Gogh's painting \textbf{\textit{The Starry Night}} satisfy classical turbulence theories or not.\citep{Aragon2008JMIV,Beattie2019arXiv,Wright2019Physics,Finlay2020JoT}


To describe turbulent flows, Lewis Fry \citet{Richardson1922} advocated a phenomenological 
picture in his seminal work "Weather Prediction by Numerical Process":
\begin{center}
\textbf{big whirls have little whirls \\
that feed on their velocity,\\
and little whirls have lesser whirls\\
and so on to viscosity\deleted{in the molecule sense}.}
\end{center}
This cascade picture has been widely accepted for describing the kinetic energy (i.e., the square of velocity) in turbulent flows qualitatively, which is transferred from large-scale to small-scale flow structures and known as the forward energy cascade.\citep{Frisch1995,Alexakis2018PR,Zhou2021PR} Later in 1941, A.N. Kolmogorov proposed his famous theory of locally homogeneous and isotropic turbulence to quantitatively characterize the Richardson's picture. According to Kolmogorov's theory, the Fourier power spectrum of kinetic energy $E(k)$ in fully-developed turbulence follows a scaling law in the so-called inertial range $k_L \ll k\ll k_{\eta}$ as,
\begin{equation}
    E(k)\propto \epsilon^{2/3} k^{-5/3},\label{eq:K41}
\end{equation}
where $\epsilon$ is the mean energy dissipation rate in units of kinetic energy per unit mass and unit time; the wavenumber $k$ is the inverse of the length scale, and the subscripts $L$ and $\eta$ indicate the system and the Kolmogorov length scales, respectively.\citep{Kolmogorov1941a} This theory, now recognized as the cornerstone in the field of turbulence, is the first theory to provide a quantitative prediction of turbulent flows and has been widely verified both experimentally and numerically.\citep{Frisch1995,Pope2000,Tsinober2009book} The reader is referred to recent papers for a review of this topic.\citep{Alexakis2018PR,Zhou2021PR}

Note that to observe Kolmogorov's $-5/3$ law, several requirements must be satisfied. An important requirement is that there should be a sufficient scale separation, which could be characterized by the Reynolds number $\si{Re}=uL/\nu$. Here, $u$ is the characteristic flow velocity and $\nu$ is the kinematic viscosity of the fluid. This general definition of the $\si{Re}$ number is often interpreted as the ratio between the inertia and the viscosity forces,\citep{Tennekes1972,Frisch1995} so the Kolmogorov's $-5/3$ law has been treated as one of the most important features of high-Re-number flows dominated by inertia forces.\citep{Frisch1995,Tennekes1972,Tsinober2009book} Surprisingly, in recent years, turbulence-like phenomena have been reported for low-Re-number and even nearly-zero-Re-number flows. These flows include the so-called elastic turbulence,\citep{Groisman2000Nature} bacterial turbulence or mesoscale turbulence,\citep{Wensink2012PNAS} and lithosphere deformation,\citep{Jian2019PRE} to list a few. In these systems, despite their small Re numbers (in the range $\mathcal{O}(10^{-24})\lesssim \si{Re}\lesssim \mathcal{O}(10^{-1})$), a wide scale separation can be still observed in the flow patterns, and thus a turbulence-like scaling behavior emerges. These findings imply that even for barely-flowing systems, one may examine their turbulence-like patterns in the framework of turbulence theories. 

 \begin{figure*}[!htb]
\centering
 \includegraphics[width=0.75\linewidth]{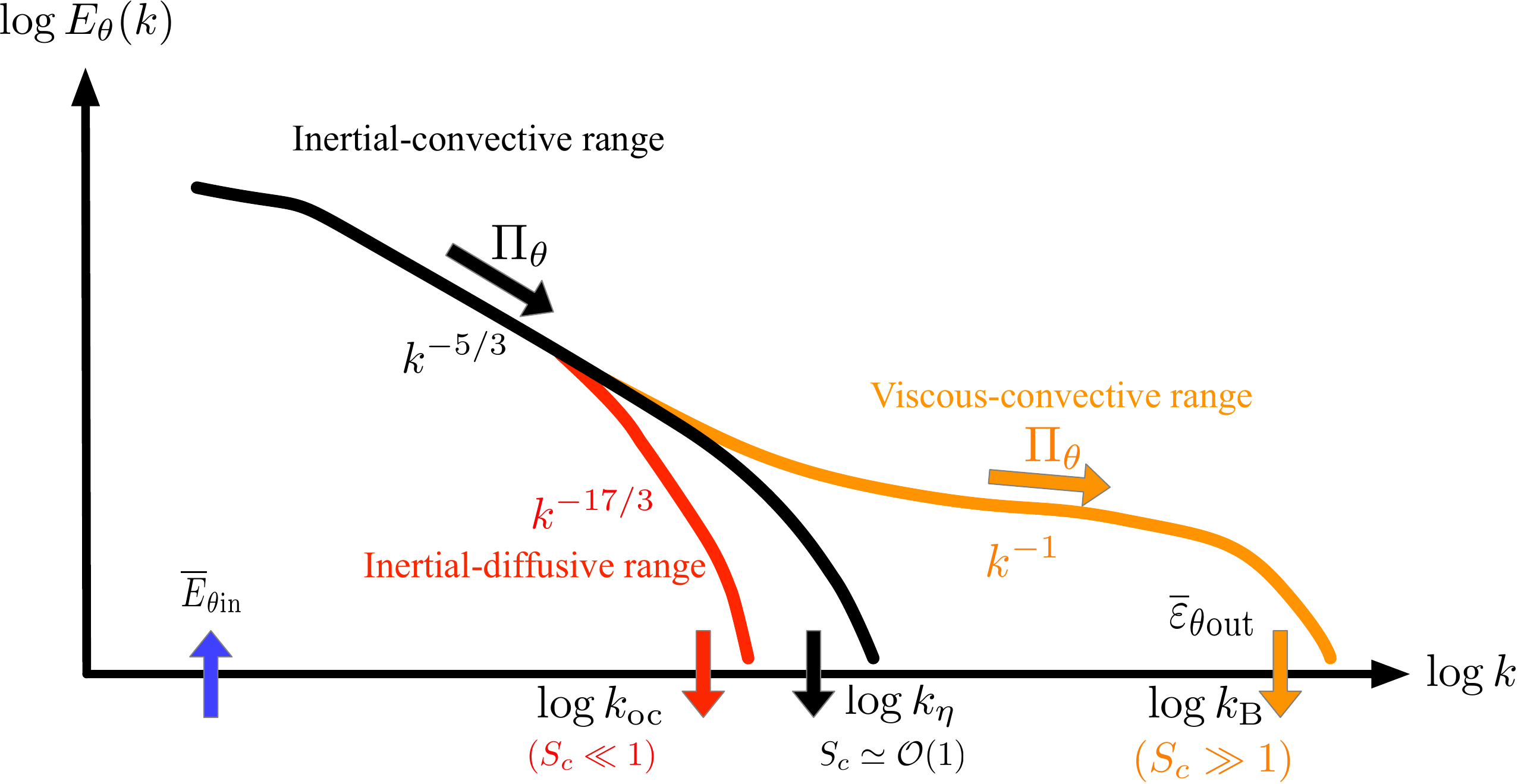}
 \caption{ Scalar spectra $E_{\theta}(k)$ for different Schmidt numbers $Sc$ reproduced from the Ref.\,\onlinecite{Sreenivasan2019PNAS}. For $Sc\gg 1$, the so-called Batchelor spectrum $E_{\theta}(k)\propto k^{-1}$ is expected to be in the range $k_{\eta}\ll k \ll k_B$, where $k_{\eta}$ and $k_{B}$ are the Kolmogorov and the Batchelor wavenumbers, respectively. See the text for a detailed explanation. 
 }\label{fig:Batechelor1959}
\end{figure*}

For art paintings, their patterns can be treated as snapshots of flow fields. However, one cannot obtain the kinetic energy information from these patterns. Instead, a more suitable quantity to characterize their features is luminance, which is a passive scalar similar as dye and temperature that are transported and mixed by the flow, so its spatial distribution is highly correlated to the characteristics of the velocity field. Quantitatively, the behavior of a passive scalar $\theta$ is determined by the Schmidt number $\si{Sc}=\nu/\kappa$, a ratio of the fluid viscosity $\nu$ to the scalar diffusivity $\kappa$.\citep{Tennekes1972,Pope2000} In terms of turbulent small-scale properties, the $\si{Sc}$ number can also be expressed using the ratio between the Batchelor wavenumber $k_{B} = (\epsilon/\nu \kappa^2)^{1/4}$ of the passive scalar and the Kolmogorov wavenumber $k_\eta = (\epsilon/\nu^3)^{1/4}$ of the velocity field: $\si{Sc}=(k_{B} / k_\eta)^2$. Depending on the value of the $\si{Sc}$ number, there exist three distinct regimes in the Fourier power spectrum of passive scalar $E_{\theta}(k)$ as illustrated in Fig.\,\ref{fig:Batechelor1959}. For $\si{Sc}=\mathcal{O}(1)$ with $k_B\simeq k_{\eta}$, a scaling behavior similar as the Kolmogorov's $-5/3$ law can be expected in the inertial-convective subrange $k_L\ll k \ll k_\eta$, i.e.,
\begin{equation}
    E_\theta (k) = C_{\si{OC}} \epsilon_{\theta} \epsilon^{-1/3} k^{-5/3},
\end{equation}
where $C_{\si{OC}}$ is the Obukhov-Corrsin constant and $\epsilon_{\theta}$ is the mean scalar dissipation rate. This is the so-called Kolmogorov-Obukhov-Corrsin scaling (KOC for short).\cite{Obukhov1949,Corrsin1951,Warhaft2000,Sreenivasan2019PNAS}
For the case with $\si{Sc}\ll 1$, one still expects the $-5/3$ scaling, but the inertial-convective subrange is shorter than that in the KOC case since $k_B< k_{\eta}$.


For the case of $\si{Sc}\gg 1$, \citet{Batchelor1959JFM} obtained the following spectrum for the scales beyond the inertial-convective subrange,
\begin{equation}
E_\theta (k) = C_{\si{B}}\epsilon_{\theta} (\nu / \epsilon)^{1/2} k^{-1}\exp\left(- C_{\si{B}} (k/k_B)^2\right), \, k\gg k_{\eta}\label{eq:Bactelor_expontential}
\end{equation}
where $C_{\si{B}}$ is the Batchelor constant. This shows that if $k_B \ll k$ (i.e., in the viscous-diffusive subrange), the spectrum follows a rapid exponential decay.\citep{Donzis2010FTC,Sreenivasan2019PNAS}
Note that in the viscous-convective subrange, i.e., $k_\eta\ll k\ll k_B$, an asymptotic power-law is expected, 
\begin{equation}
    E_\theta (k) = C_{B} \epsilon_{\theta} (\nu / \epsilon)^{1/2} k^{-1}, \label{eq:Bactelor_Scaling}
\end{equation}
Several attempts have been performed to verify the Batchelor's $-1$ scaling either experimentally or numerically and the evidence has become increasingly convincing in recent years.\citep{Gibson1963JFM,Wu1995PRL,Antonia2003AMR,Yeung2004FTC,Amarouchene2004PRL,Goetzfried2019PRF,Mohaghar2020PRF,Iwano2021EiF,Bedrossian2022CPAM,Saito2024PRF} However, due to the the lack of a clear scale separation, it remains challenging to observe both the KOC's $-5/3$ scaling and the Batchelor's $-1$ scaling simultaneously, which requires at least 3$\sim$4 orders of scale separation in experiments or numerical simulations to resolve all dynamically relevant scales.\citep{Sreenivasan2019PNAS}

Concerning the \textbf{\textit{The Starry Night}} examined in the present study, it was painted by linseed oil (high fluid viscosity) mixed with stone powder (low scalar diffusivity), implying a high Sc number. Therefore, one might be curious about whether the flow pattern in this artwork adheres to the Batchelor's theory of scalar turbulence. \citet{Aragon2008JMIV} found that the increment of the luminance in this painting shows a clear scale invariance, and the corresponding probability density functions can be reproduced using the formula obtained from the turbulence theory. \citet{Beattie2019arXiv} showed that the Fourier power spectrum of the luminance is close to $E_{\theta}(k)\propto k^{-2}$ rather than the Kolmogorov $-5/3$ scaling, which could be interpreted using the theory of compressible turbulence. However, \citet{Finlay2020JoT} reported that the midrange wavenumber spectrum tends to obey a $-1$ scaling. These results seem to contradict each other, partially because their examined areas of the painting were not exactly the same, so the spectrum might be contaminated by different elements in the painting. Moreover, these studies considered only part of the painting and thus some whirls, which are crucial for characterizing the multi-scale feature of turbulence, were excluded in their analysis; see Fig.\,\ref{fig:StarryNight}\,(b).

In this work, we revisit the controversial issue above by keeping {\textit{all and only}} the whirls in \textbf{\textit{The Starry Night}} during the analysis, following the fundamental hypothesis of Richardson-Kolmogorov's cascade picture of turbulence. Both the Fourier power spectrum and the second-order structure function \added{of the gray-scale luminance of the painting} are analyzed. Their scaling behaviors are then compared with the prediction of the Batchelor theory of scalar turbulence. The implication of our findings will be discussed.

\section{Data and Method}

\subsection{High Resolution Version of \textbf{\textit{The Starry Night}}}\label{sec:data}


\begin{figure*}[!htb]
\centering
 \includegraphics[width=1\linewidth]{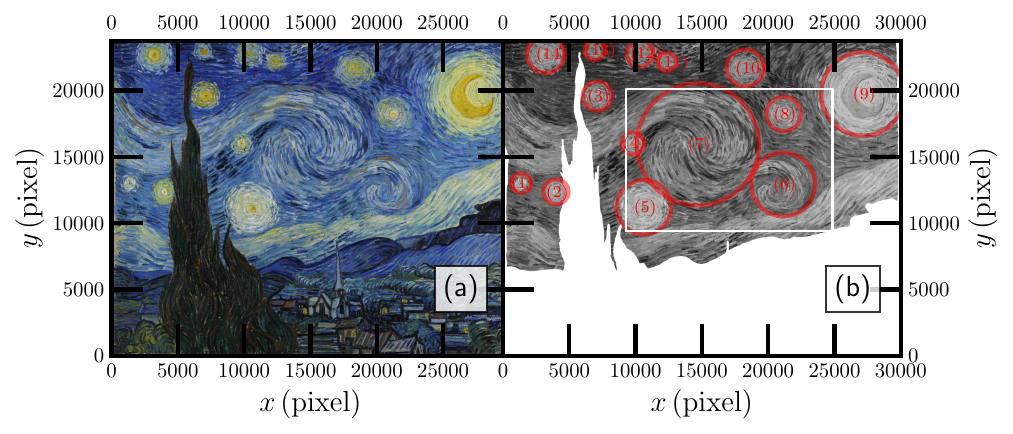}
 \caption{(Color online) (a) a high-resolution van Gogh's \textbf{\textit{The Starry Night}} obtained from \href{https://artsandculture.google.com/story/egVRmbCQ5tyrVA}{https://artsandculture.google.com} with a size $92.1 \si{cm} \times 73.7 \si{cm}$ and $30,000\,\si{pixel}\times 23,756\,\si{pixel}$. Visually, the sky seems to be flowing with swirling eddies. (b) Gray version of the \textbf{\textit{The Starry Night}}, where the region studied by \citet{Finlay2020JoT} is illustrated by a white square. The non-flow part is masked out manually. The whirls/eddies are recognized by naked eyes.}\label{fig:StarryNight}
\end{figure*}

 \textbf{\textit{The Starry Night}} is an oil-on-canvas painting by the Dutch postimpressionist painter Vincent van Gogh painted in June 1889. It depicts the view from the east-facing window of his asylum room at Saint-R\'emy-de-Provence, south of France, just before sunrise, with the addition of an imaginary village and flowing sky; see Fig.\,\ref{fig:StarryNight}. It has been in the permanent collection of the Museum of Modern Art in New York City since 1941, acquired through the Lillie P. Bliss Bequest. \textbf{\textit{The Starry Night}}, widely regarded as Vincent van Gogh's magnum opus, is one of the most recognized paintings in western art and can be widely found in our daily life; see Fig.\,\ref{fig:StarryNight_daily_Appendix} in the Appendix.
 
 Fig.\,\ref{fig:StarryNight} shows a high-resolution version of \textbf{\textit{The Starry Night}} provided by Google Art Project \href{https://artsandculture.google.com/story/egVRmbCQ5tyrVA}{(https://artsandculture.google.com).} It has a size of $92.1 \si{cm} \times 73.7 \si{cm}$ and $30,000\,\si{pixel}\times\,23,756\,\si{pixel}$, corresponding to a spatial resolution of $30\,\si{\mu m/pixel}$. Fourteen eddies (including the moon) of different sizes can be recognized by naked eyes with their diameters in the range $4.2\,\si{cm} \lesssim r \lesssim 27.6 \,\si{cm}$ (i.e., $1,400\,\si{pixel}\lesssim r \lesssim 9,200\,\si{pixel}$); see Tab.\,\ref{tab:eddy} in the Appendix. The typical spatial scale of the brushstroke is found to be in the range $0.09\,\si{cm} \lesssim r \lesssim 1.5\,\si{cm}$ (i.e., $30\,\si{pixel} \lesssim r \lesssim 500\,\si{pixel}$) for the width and $1.2\,\si{cm}\,\lesssim r \lesssim 6\,\si{cm}$ (i.e, $400\,\si{pixel}\,\lesssim r \lesssim 2,000\,\si{pixel}$) for the length; see Fig.\,\ref{fig:StarryNight_Brushstrokes_Appendix} in the Appendix.
 
 Before making the analysis, the original image is converted from the red-green-blue scale to the gray-scale using the following formula,
 \begin{equation}
   {  Y=0.2125R+0.7154G+0.0721B}
 \end{equation}
 where $R$, $G$, and $B$ represent the intensity for each color channel. The function color.rgb2gray from the Python scikit-image package is utilized for this transformation, which can well preserve the flow structures.\footnote{We conducted the same analysis for each individual channel (not shown here). Apart from the blue channel, the Fourier power spectra of the red and green channels exhibited the same $-5/3$ and $-1$ scalings as found from the gray-scale field, indicating that the flow-like structures are well maintained.} In addition, the church, mountain, and village are masked out to exclude the potential influence of these non-flow-like elements; see Fig.\,\ref{fig:StarryNight}\,(b). The so-obtained gray-scale field is subsequently treated as a passive-scalar field for the following analysis.
   
\subsection{Methods}

\subsubsection{Fourier Power Spectrum}\label{sec:method_Fourier}
As mentioned in the Introduction, when the flow is turbulent, a power-law behavior is expected for the Fourier power spectra of both the velocity and the passive scalar advected by the velocity field. Classically, the Fourier power spectrum is estimated using the Fast Fourier Transform algorithm, with datasets with a size of the form $2^p$, where $p$ is an integer. This algorithm also requires datasets with no missing values. However, the masked-out data in this work, as seen in Fig.\,\ref{fig:StarryNight}\,(b), have missing parts. In order to overcome these limitations, the Fourier power spectrum is estimated via the Wiener-Khinchine theorem here. This theorem states that, \added{for the luminance $\theta$ (e.g., the gray-scale field $Y$ defined above)}, its Fourier power spectrum \replaced{$E_{\theta}(k)$}{$E(k)$} and the autocorrelation function \replaced{$\rho_{\theta}(r)$}{$\rho(r)$} are a Fourier transform pair, which are written as, 
 \begin{equation}
\replaced{E_{\theta}(k)=\int \rho_{\theta}(r) \exp(-j 2\pi k r ) \upd r,\,\rho_{\theta}(r)=\int E_{\theta}(k) \exp(j 2\pi k r )\upd k}{E(k)=\int \rho(r) \exp(-j 2\pi k r ) \upd r,\,\rho(r)=\int E(k) \exp(j 2\pi k r )\upd k},
  \end{equation}
where $j=\sqrt{-1}$ is a complex unit; $k=1/r$ is the wavenumber and $r$ is the distance between two points in the physical space. \replaced{The autocorrelation function is defined as $\rho_{\theta}(r)=\langle \theta'(x+r)\theta'(x)\rangle $, in which $\theta'(x)=\theta(x)-\langle \theta \rangle $ is the scalar variation in space and $\langle \cdot \rangle $ means ensemble average.}{The autocorrelation function} $\rho_{\theta}(r)$ can be estimated when there are missing data, 
and in such case an additional step is involved to correct the missing data effect; see detail of this algorithm in Ref.\,\onlinecite{Gao2021JGR}. In case of scale invariance, one expects a power-law behavior of $E_{\theta}(k)$ written as below,
\begin{equation}
\replaced{E_{\theta}(k)\propto k^{-\beta_{\theta}}}{E(k)\propto k^{-\beta}},
\end{equation}
where $\beta_{\theta}>0$ is the scaling exponent that can be determined experimentally or through theoretical considerations; for example $\beta=5/3$ for the velocity spectrum of high Reynolds number flows.\citep{Kolmogorov1941a,Frisch1995,Schmitt2016Book} 

\subsubsection{Second-order Structure Function}\label{sec:method_SF}
To characterize the scale invariance in the physical space, the second-order structure-function is often used. For luminance $\theta$ examined here, this function is written as, 
\begin{equation}
\replaced{S_{\theta 2}(r)=\langle \Delta_r \theta(x)^2 \rangle\propto r^{\zeta_{\theta}(2)} }{S_{2}(r)=\langle \Delta_r \theta(x)^2 \rangle\propto r^{\zeta(2)} }\label{eq:SF2}
\end{equation}
where $\Delta_r \theta(x)=\theta(x+r)-\theta(x)$ is the scalar increment over a distance $r$; \replaced{$\zeta_{\theta}(2)$}{$\zeta(2)$} is the second-order scaling exponent if the power-law behavior holds. A scaling relation \replaced{$\beta_{\theta}=1+\zeta_{\theta}(2)$}{$\beta=1+\zeta(2)$} is expected for \replaced{$1<\beta_{\theta}<3$}{$1<\beta<3$}.\citep{Frisch1995,Schmitt2016Book}
 However, as discussed by \citet{Huang2010PRE,Huang2013PRE}, due to several reasons, for instance, contamination by the energetic large-scale structures (e.g., ramp-cliff structures in scalar turbulence\citep{Warhaft2000,Huang2011PRE}), ultraviolet or infrared effects, to name a few, this scaling relation is often violated;\citep{Warhaft2000,Huang2010PRE,Huang2013PRE} see more discussion in Ref.\,\onlinecite{Schmitt2016Book}. Note that when $\si{Sc}\gg 1$, Batchelor's theory of scalar turbulence predicts a scaling value of \replaced{$\beta_{\theta}=1$}{$\beta=1$}, and the power-law in Eq.\,\eqref{eq:SF2} is then violated due to the ultraviolet effect. For this situation, Batchelor's theory predicts a log-law, which is written as, \begin{equation}
\replaced{S_{\theta 2}(r)\propto \alpha_{\theta} \ln(r)}{S_{2}(r)\propto \alpha \ln(r)} \label{eq:loglaw}
 \end{equation}
where $r_{B} \ll r\ll r_{\eta}$ and $\alpha_{\theta}$ is an unknown parameter. Therefore, instead of the power-law in Eq.\,\eqref{eq:SF2}, the log-law in Eq.\,\eqref{eq:loglaw} will be tested in the present study.

\section{Results}

\subsection{Fourier Power Spectrum}

\begin{figure}[!htb]
\centering
 \includegraphics[width=0.8\linewidth]{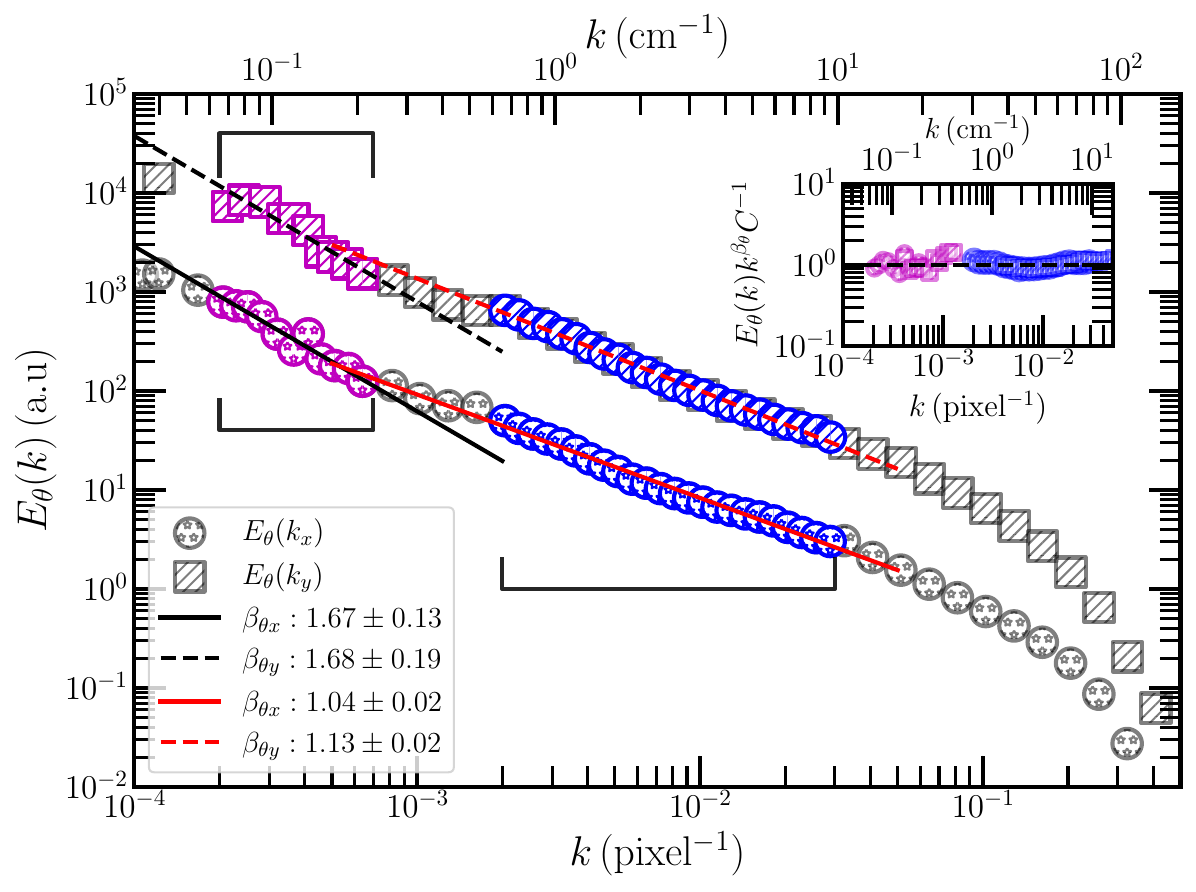}
 \caption{(Color online) Experimental Fourier power spectrum $E_{\theta}(k)$, where the black and red lines indicate the power-law behaviors in the ranges $6.67\times 10^{-2}\,\si{cm^{-1}}\lesssim k \lesssim 2.33\times 10^{-1}\,\si{cm^{-1}}$ (i.e.,
 $2\times 10^{-4}\,\si{pixel}^{-1}\,\lesssim k\lesssim 7\times 10^{-4}\,\si{pixel}^{-1})$ and $6.67 \times 10^{-1}\,\si{cm^{-1}}\,\lesssim k \lesssim 10\,\si{cm^{-1}}$ (i.e., $2\times 10^{-3} \,\si{pixel^{-1}} \lesssim k \lesssim 3\times 10^{-2} \,\si{pixel^{-1}}$), respectively. For clarity, the curve $E_{\theta}(k_y)$ has been shifted up by multiplying a factor of 10.
 The inset shows the compensated curves $E_{\theta}(k)k^{\beta_{\theta}}C^{-1}$ using the corresponding scaling exponents $\beta_{\theta}$ and prefactors $C$ to emphasize the power-law behaviors.}\label{fig:StarryNight_Spectrum}
\end{figure}

\begin{figure}[!htb]
\centering
 \includegraphics[width=0.9\linewidth]{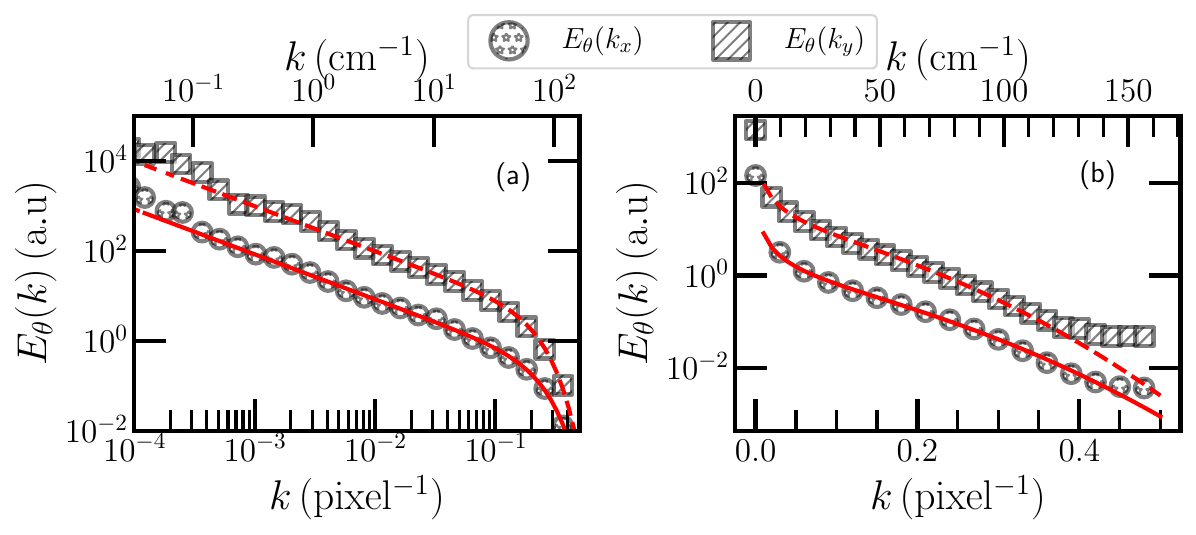}
 \caption{(Color online) Experimental verification of Eq.\,\eqref{eq:Bactelor_expontential}, where the solid and dashed lines are least squares fits to the data in the range $6.67\times 10^{-1}\,\si{cm^{-1}}\lesssim k \lesssim 1.33\times 10^2\,\si{cm^{-1}}$ (i.e., $2\times 10^{-3}\,\si{pixel}^{-1}\lesssim k \lesssim 4\times 10^{-1}\,\si{pixel}^{-1}$) for $E_{\theta}(k_x)$ and $E_{\theta}(k_y)$, respectively: (a) a log-log plot to highlight the power-law behavior $E_{\theta}(k) \sim k^{-1}$; (b) a semilog-y plot to highlight the exponential tail $E_{\theta}(k) \sim \exp\left( -\left(k/k_B \right)^2\right)$. For clarity, the curve $E_{\theta}(k_y)$ has been shifted up by a multiplying a factor of 10.}\label{fig:StarryNight_Batchelor}
\end{figure}

The Fourier power spectra $E_{\theta}(k)$ are estimated along the horizontal ($x$) and vertical ($y$) directions using the algorithm described in Sec.\,\ref{sec:method_Fourier}. A bin average with 10 points per order of wavenumber is performed. Fig.\,\ref{fig:StarryNight_Spectrum} shows the thus-obtained $E_{\theta}(k)$, where a dual power-law behavior is visible. As mentioned in Sec.\,\ref{sec:data}, the spatial sizes of the whirls are in the range $4.2\,\si{cm}\lesssim r \lesssim 27.6\,\si{cm}$ (i.e., $1,400\, \si{pixel}\lesssim r \lesssim 9,200\,\si{pixel}$), we therefore attempt power-law fit to the data in this range, following the Richardson-Kolmogorov's cascade picture of turbulence. It is found that power-law behaviors can be well determined in the wavenumber range $6.67\,\times 10^{-2} \,\si{cm^{-1}}\,\lesssim k \lesssim 2.33 \times 10^{-1}\,\si{cm^{-1}}$ (i.e., $2\times 10^{-4}\,\si{pixel^{-1}}\lesssim k \lesssim 7\times 10^{-4}\,\si{pixel^{-1}}$), corresponding to the spatial scale in the range $4.3\,\si{cm}\lesssim r \lesssim 15\,\si{cm}$ (i.e., $1,430\,\si{pixel}\lesssim r\lesssim 5,000\,\si{pixel}$). The scaling exponents are found to be $\beta_{\theta x}=1.67\pm 0.13$ and $\beta_{\theta y}=1.68\pm0.19$, where the $95\%$ fit confidence is provided by the least squares fit algorithm. These values agree well with the one predicted by the KOC theory, since the scaling range chosen here satisfies the requirement of the Richardson-Kolmogorov's cascade picture of turbulence, where the whirls/eddies that cover a sufficient scale range are included in the analysis.\citep{Richardson1922,Kolmogorov1941a,Frisch1995,Schmitt2016Book} \added{This finding implies that the arrangement of the eddy-like formations crafted by van Gogh resembles the energy transfer mechanism in real turbulent flows.}

The second power-law behavior is observed in the wavenumber range $6.67\times 10^{-1}\,\si{cm^{-1}}\lesssim k \lesssim 10\,\si{cm^{-1}}$ (i.e., $2\times 10^{-3}\,\si{pixel^{-1}} \lesssim k \lesssim 3\times 10^{-2}\,\si{pixel^{-1}}$), corresponding to the spatial scale in the range $ 0.1\,\si{cm}\lesssim r\lesssim 1.5\,\si{cm}$ (i.e., $33\,\si{pixel}\lesssim r \lesssim 500\,\si{pixel}$). The measured scaling exponents are found to be $\beta_{\theta x}=1.04\pm 0.02$ and $\beta_{\theta y}=1.13\pm0.02$, close to the Batchelor $-1$ scaling. \added{As we discussed in the Introduction, such a scaling is expected to observe in the viscous-convective range of scalar turbulence.\citep{Yeung2004FTC,Clay2017PhD,Sreenivasan2019PNAS}  Notably, the wavenumber range for the $-1$ scaling is in line with that of the brushstroke width, suggesting that the diffusion and mixing properties associated with the painting process may result in patterns that resemble the diffusion and mixing observed in turbulent flows.} 
 
To highlight the two distinct power-law behaviors, the compensated curves using the fitted parameters are shown in Fig.\,\ref{fig:StarryNight_Spectrum} as inset, where clear plateaus are observed. From Fig.\,\ref{fig:StarryNight_Spectrum}, one can also observe a fast decay of $E_{\theta}(k)$ in the large wavenumber range, motivating us to check Eq.\,\eqref{eq:Bactelor_expontential} predicted by  Batchelor.\citep{Batchelor1959JFM} To do so, the least squares fit algorithm is performed to the curve $E_{\theta}(k)$ in the range $6.67\times 10^{-1}\,\si{cm^{-1}}\lesssim k \lesssim 1.33\times 10^2\,\si{cm^{-1}}$ (i.e., $2\times 10^{-3}\,\si{pixel^{-1}}\lesssim k \lesssim4\times 10^{-1}\,\si{pixel^{-1}}$). Visually, Eq.\,\eqref{eq:Bactelor_expontential} fits the data well with a Batchelor-like parameter $k_B=67\pm6\,\si{cm^{-1}}$, corresponding to a spatial scale of $0.015\,\si{cm}$ (5 pixel); see Fig.\,\ref{fig:StarryNight_Batchelor}\,(a). To highlight the exponential tail $E_{\theta}(k) \sim \exp\left( -\left(k/k_B \right)^2\right)$, the results are re-plotted in a semilog-y view; see Fig.\,\ref{fig:StarryNight_Batchelor}\,(b), which confirms the validation of Eq.\,\eqref{eq:Bactelor_expontential}. 

\subsection{Second-order Structure Function}

\begin{figure}[!htb]
\centering
 \includegraphics[width=0.8\linewidth]{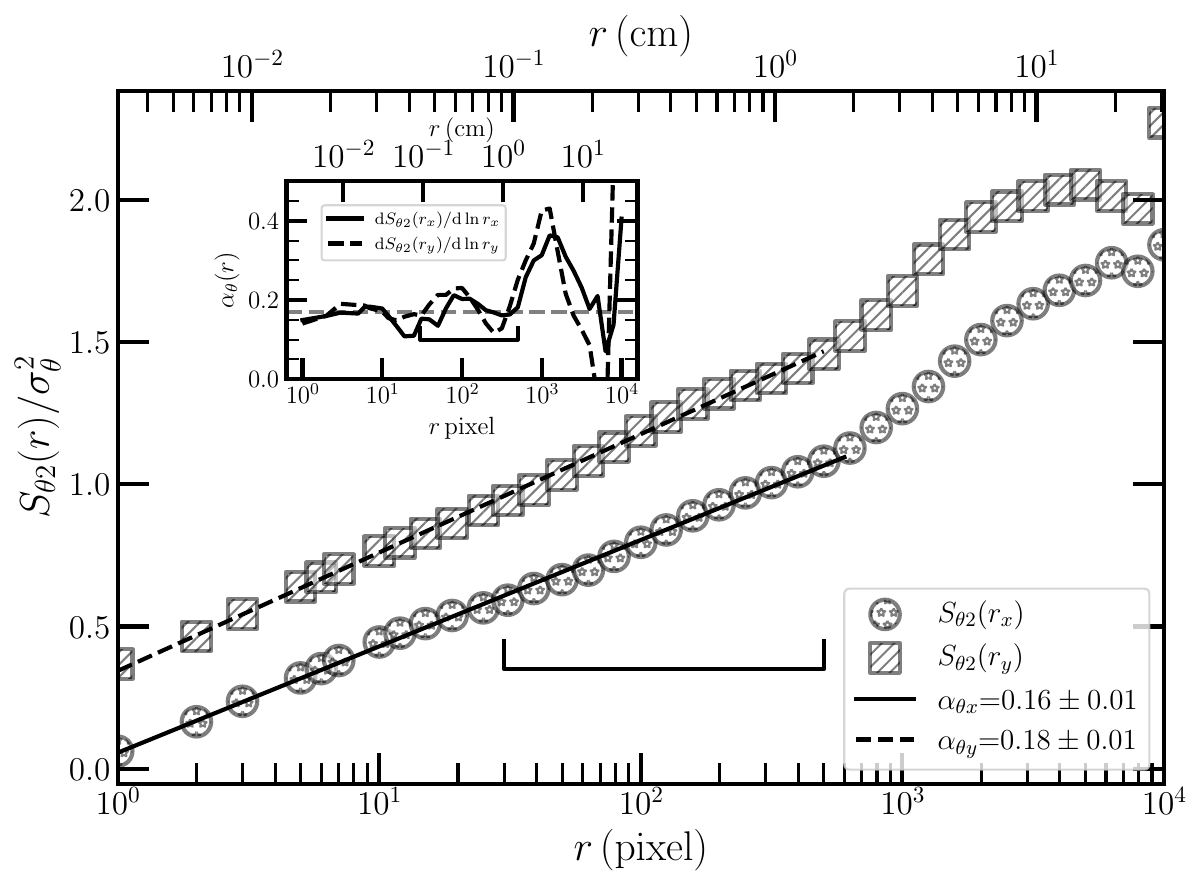}
 \caption{(Color online) Experimental verification of Eq.\,\eqref{eq:loglaw} in a semilog-x plot, where the solid and dashed lines are least squares fit to the data in the range $ 0.003\,\si{cm}\lesssim r\lesssim 1.5\,\si{cm}$ (i.e., $1\,\si{pixel}\lesssim r\lesssim 500\,\si{pixel}$) for $S_{\theta 2}(r_x)$ and $S_{\theta 2}(r_y)$, respectively. For display clarity, the curve of $S_{\theta 2}(r_y)$ has been shifted up vertically by adding a constant of $0.3$. The inset shows the local slope $\alpha_{\theta}(r) = \upd \left(S_{\theta 2}(r)/\sigma_{\theta}^2\right)/\upd \ln (r)$, where the horizontal dash line indicates a mean value of $\overline{\alpha}_{\theta} = 0.17\pm0.03$.}\label{fig:StarryNight_SF}
\end{figure}

As mentioned in Sec.\,\ref{sec:method_SF}, the power-law behavior of the second-order structure function might be strongly biased due to the presence of the ultraviolet effect (e.g., the observation of $\beta_{\theta}\simeq 1$) in the present study. Therefore, instead of Eq.\,\eqref{eq:SF2}, the log-law in Eq.\,\eqref{eq:loglaw} is examined. Fig.\,\ref{fig:StarryNight_SF} shows the estimated second-order structure functions $S_{\theta 2}(r)$ normalized by the luminance variance of the examined region of the painting. A clear logarithmic law is evident in the range $ 0.003\,\si{cm}\lesssim r\lesssim 1.5\,\si{cm}$ (i.e., $1\,$pixel$\lesssim r \lesssim 500$\,pixel), with the fitting slopes being $0.16\pm0.01$ and $0.18\pm0.01$ for the horizontal and vertical directions, respectively. Note that this log-law range is compatible with the range of the brushstroke width. The local slope $\alpha_{\theta}(r) = \upd \left(S_{\theta 2}(r)/\sigma_{\theta}^2\right)/\upd \ln (r)$ is also estimated using a finite center difference; see the inset in Fig.\,\ref{fig:StarryNight_SF}. Generally speaking, $\alpha_{\theta}(r_x)$ and $\alpha_{\theta}(r_y)$ have the same evolution trend, with a mean value of $\overline{\alpha}_{\theta}=0.17\pm0.03$ in the range mentioned above. Combined with the findings in the Fourier power spectrum, it seems that Batchelor's scalar turbulence theory is a good candidate for interpreting the present results phenomenologically. 

 
\section{Discussions}

\subsection{Turbulent Flows in Art Paintings}

Science and art often inspire each other. To what degree the complex physics of natural flows can be captured by the patterns in artworks has attracted growing interest from the community of fluid dynamics.
For example, using a physics-informed deep learning framework that is capable of encoding
the Navier-Stokes equations into neural networks, \citet{Raissi2020Science} successfully extracted the velocity and pressure fields from Leonardo da Vinci's painting of turbulent flows. 
\citet{Colagrossi2021PoF} reproduced the physics behind one of Leonardo da Vinci's drawings (i.e., a water jet impacts on a pool painted in 1510-1512) by a smoothed particle hydrodynamic model, and concluded that Leonardo da Vinci ``was able to extract essential phenomena of complex air-water flows and accurately describe each flow feature independently of the others, both in his drawings and in their accompanying notes''. In fact, Leonardo da Vinci is considered one of the pioneers in identifying the characteristic feature of turbulent flows, as evidenced by the multi-scale eddies pattern depicted in several of his artworks.\citep{Chen2019ME,Marusic2021ARFM,Warhaft2022AS}


Concerning \textbf{\textit{The Starry Night}} painted by Vincent van Gogh, our results show a clear evidence of the $-5/3$ scaling law when all and only the whirls/eddies in the painting are included in the analysis. According to the Richardson-Kolmogorov's cascade picture of turbulence, a sufficient number of eddies with a wide distribution of scales should be involved to observe the $-5/3$ scaling; see more examples in the Appendix\,\ref{sec:SI_Images}. Our present finding thus suggests that, not only the size distribution of whirls/eddies in \textbf{\textit{The Starry Night}} but also their relative distance and intensity follow the physical law that governs the behaviors of turbulent flows. In other words, Vincent van Gogh had a very careful observation of real flows, and the $-5/3$ scaling observed here is due to this excellent mimic of real flows.

\subsection{Estimation of the Reynolds and the Schmidt numbers}

\added{As previously noted, the Richardson-Kolmogorov $-5/3$ scaling requires a wide range of scales, usually associated with high Reynolds number flows. The $-5/3$ scaling revealed here arises from the artist's representation of real flows, as opposed to the nonlinear interactions between multi-scale eddies in hydrodynamic turbulence. Meanwhile, the $-1$ scaling could result from physical processes like diffusion and mixing during painting. }
According to the Batchelor's theory of scalar turbulence, one should have a stationary flow with the Schmidt number $\si{Sc}\gg 1$ to observe the -1 scaling.\citep{Batchelor1959JFM} The former condition is automatically satisfied, since the flows during preparing the painting oil and the painting process are slow enough. \added{The latter condition is arguably satisfied, as \textbf{\textit{The Starry Night}} was painted by linseed oil (high fluid viscosity) mixed with stone powder (low scalar diffusivity). To check these conditions quantitatively,} we estimate the Reynolds and the Schmidt numbers as follows.

As mentioned in Sec.\,\ref{sec:data}, the length of the brushstroke is in the range $1.2\,\si{cm}\lesssim r \lesssim 6\,\si{cm}$.
We therefore take the median value, that is $L=3.6\,$cm, as the characteristic length scale.
Assuming that the typical time scale for each brushstroke is $1\,\si{sec}$, then the typical velocity during the painting is around $u\simeq 3.6\,\si{cm/s}$. Therefore, the Reynolds number is estimated to be $\si{Re}=uL/\nu_{\si{eff}}\simeq 19.1 \propto \mathcal{O}(10)$, 
where $\nu_{\si{eff}}\simeq 6.79 \times 10^{-5}\si{ m^2/s}$ is the effective kinematic viscosity estimated by the Einstein relation approximately; see Appendix\,\ref{sec:thermal} for the estimation in detail.\citep{Einstein1905AP}


Note that the Reynolds number can also be expressed as the separation ratio of the characteristic scales in turbulent flows,\citep{Pope2000} i.e.,
\begin{equation}
    \si{Re} \propto \left(\frac{L_E}{\eta_k}\right)^{4/3}
\end{equation}
where $L_E$ represents the size of the largest eddy and $\eta_k$ is the Kolmogorov dissipation scale. In the present study, we can estimate the value of $L_E$ from the painting, being $L_E \simeq 27.6\,\si{cm}$ approximately. For the value of $\eta_k$, Fig.\,\ref{fig:StarryNight_Spectrum} shows that the $-5/3$ scaling and the Batchelor-like scaling are observed in the spatial scale ranges $4.3 \,\si{cm} \lesssim r \lesssim 15\,\si{cm}$ and $0.1\,\si{cm} \lesssim r \lesssim 1.5\,\si{cm}$, respectively; so $\eta_k$ should lie between ${1.5}\,\si{cm}$ and ${4.3}\,\si{cm}$. Then the Reynolds number estimated from Eq.(10) is in the range $11.9 \lesssim \si{Re} \lesssim  48.6$, which is also $\mathcal{O}(10)$ and consistent with the value estimated above.
\added{The Taylor microscale Reynolds number can be further calculated using the well-known formula $\si{Re}_{\lambda}=\left(\frac{20}{3}\si{Re}\right)^{1/2}$ \citep{Pope2000}, resulting in a range $9\lesssim \si{Re}_{\lambda}\lesssim 18$.}

As for the Schmidt number, its value can be calculated by $\si{Sc}=(k_{B} / k_\eta)^2 = \left(\eta_k/\eta_B\right)^2$. The Batchelor-like scale $\eta_B$ has been obtained from Fig.\,\ref{fig:StarryNight_Batchelor}, which is $\eta_B\simeq 0.015\,\si{cm}$. And since $1.5 \,\si{cm} \lesssim \eta_k \lesssim 4.3\,\si{cm}$ as discussed above, the low bound of the Schmidt number is estimated to be $\si{Sc} \simeq \left(1.5/0.015\right)^2=\mathcal{O}(10^4)$. \added{Alternatively, the Schmidt number can be approximated by using its original definition: $\si{Sc}=\nu_{\si{eff}}/\kappa_{\si{eff}}=\mathcal{O}(10^{11})$, where $\nu_{\si{eff}}\simeq 6.79 \times 10^{-5}\si{m^2/s}$ and $\kappa_{\si{eff}}\simeq 3.90\times 10^{-16} \si{m^2s}$ are the effective kinematic viscosity and diffusivity coefficient estimated using the Einstein relation; refer to Appendix \ref{sec:thermal} for details. Both estimation methods yield a value of $\si{Sc}\gg 1$.} Therefore, the requirement for Batchelor's theory of scalar turbulence is satisfied.

\subsection{Batchelor Scalar Turbulence}

As mentioned in the Introduction, the prediction of the Batchelor's theory of scalar turbulence is difficult to realize not only in experiments but also in numerical simulations.\citep{Sreenivasan2019PNAS} Several attempts have been made to verify this theory. For example, \citet{Amarouchene2004PRL} observed Batchelor scaling for the thickness fluctuation of fast-flowing soap films. However, to fit the experiment spectrum curve, instead of Batchelor's original proposal $k^{-1}\exp\left(-\left(k/k_B \right)^2\right)$, an exponential tail is considered, that is, $k^{-1}\exp\left(-k/k_B\right) $, the form proposed by \citet{Kraichnan1968PoF} when the fluctuation of the strain is taken into account. Here, we can fit the experimental curve using Batchelor's original proposal, since the basic assumption of his theory of scalar turbulence is satisfied. 

\added{Numerically, \citet{Clay2017PhD} has examined the asymptotic behavior of the Batchelor's prediction via direct numerical simulations of isotropic turbulence at $\si{Re}_{\lambda} \simeq 140$ with $4 \lesssim \si{Sc} \lesssim 512$. It is found that with increasing the $\si{Sc}$ number, a wider range of scales is developed in the scalar field, resulting in a more pronounced $-1$ scaling in the Fourier power spectrum $E_{\theta}(k)$. In this context, one may anticipate that the Batchelor's $-1$ scaling could be attainable at a lower Reynolds number with a larger Schmidt number. Indeed, \citet{Yeung2004FTC} have observed a Batchelor-like scalar spectrum at $\si{Re}_{\lambda} \simeq 8$ by increasing the Sc number from 64 to 1024, which is close to the values of $\si{Re}_{\lambda}$ and $Sc$ numbers estimated in the present study and thus provides a support of our finding. }

It is important to highlight two recent notable studies of scalar turbulence.\citep{Iwano2021EiF,Saito2024PRF} \citet{Iwano2021EiF} conducted a turbulent jet experiment with a Schmidt number $\si{Sc}\simeq 3,000$ and a Reynolds number $\si{Re}_{\lambda}\simeq 200$. Dye concentration was measured at a fixed point using an optical fiber LIF (laser-induced fluorescence) probe with a spatial resolution of $2.8\,\si{\mu m}$. Utilizing Taylor's frozen hypothesis,\citep{Frisch1995,He2017ARFM} the observed six-order magnitude of wavenumber power spectra indicated the coexistence of Kolmogorov and Batchelor scalings. However, as \citet{He2017ARFM} noted, the application of Taylor's frozen hypothesis should be approached with caution. \citet{Saito2024PRF} conducted a direct numerical simulation of the passive scalar under a special setup, where the passive scalar was carried by particles in isotropic turbulence to achieve large Schmidt numbers with a Reynolds number as high as $\si{Re}_{\lambda}\simeq 500$. Their Fourier power spectra provided clear evidence of the coexistence of Kolmogorov $-5/3$ scaling  and Batchelor $-1$ scaling  over a scale range of one order of magnitude for each. Nonetheless, simultaneous observation of Kolmogorov's $-5/3$ scaling and Batchelor's $-1$ scaling through direct experimental measurements in the spatial domain remains challenging. The findings of the present study may inspire experimental approaches like ``painting in turbulent flows'' to address this issue in the future.

\section{Conclusion}

In summary, we show in this work that when all eddies in the painting is considered in the analysis, the turbulence-like statistics can be recovered for the \textbf{\textit{The Starry Night}}, with a Kolmogorov $-5/3$ scaling corresponding to the
multi-scale eddies represented by the painter, and a Batchelor $-1$ scaling produced by the oil of the painting, 
corresponding to the viscous-convective range. In other words, Vincent van Gogh, as one of the most notable post-impressionist painters, had a very careful observation of turbulent flows: he was able to reproduce not only the size of whirls/eddies, but also their relative distance and intensity in his painting. 
Furthermore, the full Batchelor spectrum (i.e., Eq.\,\eqref{eq:Bactelor_expontential}) is found for spatial scales below the size of the eddies. This is because during the preparation of the painting oil and the drawing process, the characteristic Reynolds number is low and the diffusivity is dominant. This is nicely confirmed by the second-order structure function which precisely follows the theoretical prediction, showing a log-law. This study thus reveals the hidden turbulence in the painting \textbf{\textit{The Starry Night}} using both Kolmogorov's and Batchelor's theories. 

\begin{acknowledgments}
The authors express their gratitude to Mr. Fulian Gan, Ms. Ruoyi Xie, Ms. Xuan Lei, and Ms. Xiangying Li for providing us with their photographs.
This work is sponsored by the National Natural Science Foundation (Nos. 12102165 and U22A20579).
 

\end{acknowledgments}

\section*{AUTHOR DECLARATIONS}
\section*{Conflict of Interest}
The authors have no conflicts to disclose.

\section*{Author Contributions}

\textbf{Y.X. Huang:} Conceptualization (lead); Formal analysis (lead); Investigation (lead); Writing - review \& editing (lead). 
\textbf{Y.X. Ma:} 
Formal analysis (supporting); Methodology (supporting).
\textbf{S.D. Huang:} Investigation (supporting); Writing - review \& editing (supporting). 
\textbf{W.T. Cheng:} Formal analysis (supporting); Investigation (supporting). 
\textbf{X. Lin:} Investigation (supporting); Writing - review \& editing (supporting). 
\textbf{F.G. Schmitt:} Investigation (supporting); Writing - review \& editing (supporting).

\section{DATA AVAILABILITY}
The data that support the findings of this study are available
at {\href{https://artsandculture.google.com/story/egVRmbCQ5tyrVA}{https://artsandculture.google.com}}, {\href{https://www.tate.org.uk/art/artworks/constable-chain-pier-brighton-n05957}{https://www.tate.org.uk}} and {\href{https://www.planetary.org/space-images/voyager-1-view-of-the-great-red-spot}{https://www.planetary.org}}.
A copy of the source code for the present analysis is available at {\href{https://github.com/lanlankai}{https://github.com}}.

\clearpage

\appendix
\counterwithin{figure}{section}

\section{Typical Spatial Scales }
The detection of the scaling range should follow the requirement of turbulence theories, that is, there should be enough whirling structures involved in the statistics. Here, we manually estimate the typical spatial scale for both visualized whirls and brush strokes. 

\subsection{Spatial Scales of Whirls}

The spatial sizes of fourteen whirls/eddies are estimated by naked eyes. Their diameters, locations, and areas are listed in Tab.\,\ref{tab:eddy}.
Following Richardson's picture of turbulent energy cascade, the Kolmogorov $-5/3$ scaling is expected in the range $1,400\,\si{pixel}\lesssim r \lesssim 9,200\,\si{pixel}$ (i.e., $4.2\,\si{cm}\lesssim r\lesssim 27.6\,\si{cm}$), corresponding to a wavenumber range $1\times 10^{-4}\,\si{pixel^{-1}}\lesssim k\lesssim 7\times 10^{-4}\,\si{pixel^{-1}}$ (i.e., $ 3\times 10^{-2}\,\si{cm}^{-1}\lesssim k\lesssim 2\times 10^{-1}\,\si{cm}^{-1}$). Two distinct types of structures can be visually distinguished. The first type resembles an eddy with a ring-shaped pattern, while the other one is spiral in nature; see Fig.\,\ref{fig:StarryNight_Brushstrokes_Appendix}.

\begin{table}[!htb]
\centering
\caption{Geometric properties of eddies in \textbf{\textit{The Starry Night}} manually checked by naked eyes. The diameters of the whirls/eddies are roughly in the range $1,400\,\si{pixel}\sim 9,200\,\si{pixel}$ (i.e., $4.2\,\si{cm}\sim 27.6\,\si{cm}$), corresponding to a scale ratio around $\simeq$ 6.6. The Kolmogorov-like $-5/3$ scaling law is expected in this range.}\label{tab:eddy}
\begin{tabular}{c|c|c|c|c}
\hline
No.& $D$ (pixel/$\si{c m}$)& location x (pixel) & location y (pixel)& area (pixel$^2$/$\si{c m}^2$)\\
\hline
1  & 1,500/4.5& 1,268& 12,987& 1,767,146/15.9\\
\hline
2   & 1,900/5.7& 3,926& 12,320& 2,835,287/25.5\\
\hline
3  & 2,200/6.6& 7,015& 19,662& 3,801,327/34.2\\
\hline
4  & 1,700/5.1 & 9,721& 15,973& 2,269,801/20.4\\
\hline
5  & 4,100/12.3 & 10,549& 11,114& 13,202,543/118.8\\
\hline
6  & 4,800/14.4 & 20,998& 12,857& 18,095,573/162.9\\
\hline
7  & 9,200/27.6 & 14,625& 15,861& 66,476,101/598.3\\
\hline
8  & 2,600/7.8 & 21,070& 18,235& 5,309,292/47.8\\
\hline
9  & 6,300/18.9 & 27,113& 19,718& 31,172,453/280.1\\
\hline
10 &2,800/8.4 & 18,180& 21,795& 6,157,522/55.4\\
\hline
11 &1,400/4.2 & 12,279& 22,129& 1,539,380/13.9\\
\hline
12 & 2,000/6.0 & 10,230& 22,759& 3,141,593/28.3\\
\hline
13 & 1,500/4.5 & 6,762& 23,074& 1,767,146/15.9\\
\hline
14 & 2,800/8.4& 3,076& 22,722& 6,157,522/55.4\\
\hline
\end{tabular}
\end{table}

\subsection{Spatial Scales of Brushstrokes}

\begin{figure*}
\centering
 \includegraphics[width=\linewidth]{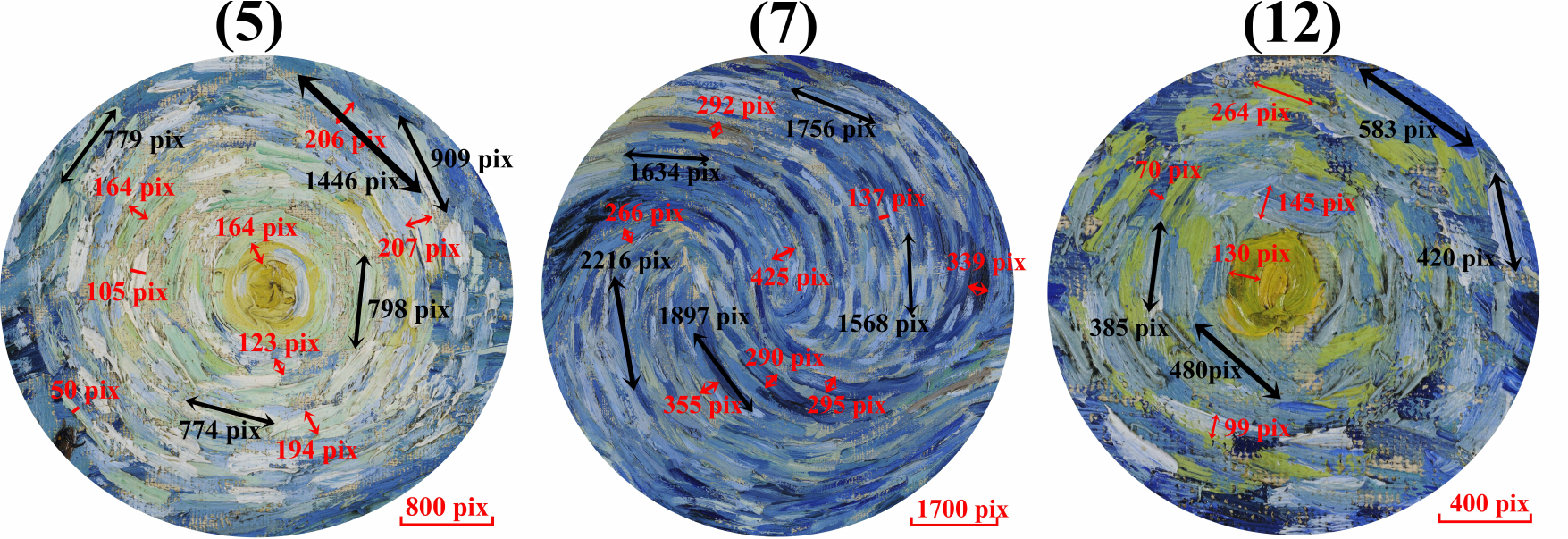}
 \caption{(Color online) Typical spatial scales of brushstrokes for the Nos.\,(5), (7) and (12) eddies marked in Fig.\,\ref{fig:StarryNight} of the main text. The width (red line) and length (black line) are found roughly in the range $30\,\si{pixel}\lesssim r \lesssim 500\,\si{pixel}$ (i.e., $0.09\,\si{cm}\lesssim r\lesssim 1.5\,\si{cm}$) and $400\,\si{pixel}\lesssim r\lesssim 2,000\,\si{pixel}$ (i.e., $1.2\,\si{cm}\lesssim r \lesssim 6\,\si{cm}$), respectively. The variation of the luminance in this range is thought to be caused by the preparation of painting oil and diffusion of the solid particles.}\label{fig:StarryNight_Brushstrokes_Appendix}
\end{figure*}

 The spatial scales of the brushstrokes are estimated manually with the width and length around $30\,\si{pixel}$ and $500\,\si{pixel}$ (i.e., $0.09\,\si{cm}\lesssim r\lesssim 1.5\,\si{cm}$) and around 400\,pixel and $2,000\,\si{pixel}$ (i.e., $1.2\,\si{cm}\lesssim r\lesssim 6\,\si{cm}$), respectively. Fig.\,\ref{fig:StarryNight_Brushstrokes_Appendix} shows an example of three typical whirls/eddies. The Batchelor's $-1$ scaling law is expected in these ranges. 
 
\clearpage

\section{The Effective Kinematic Viscosity and Diffusivity Estimated From Kinematic Dynamics}\label{sec:thermal}

\textbf{\textit{The Starry Night}} is an oil-on-canvas painting by Vincent van Gogh in 1889. At that time, the painting oil was made of stone powder and linseed oil. Using the classical knowledge of the thermal dynamics, the effective kinematic viscosity can be roughly estimated as follows.

Concerning stone powder in linseed oil, we can use a model called the Einstein equation to estimate its effective kinematic viscosity,\citep{Einstein1906AP} which is written as, 
\begin{equation}
    \mu_{\si{eff}} = \mu_{\si{f}}  (1 + 2.5\phi),\label{eq:effective_mu}
\end{equation}
where $\mu_{\si{eff}}$ is the effective dynamic viscosity of the suspension, $\mu_{\si{f}}$ is the dynamic viscosity of the fluid, and $\phi$ is the volume fraction of the particles in the suspension. It is an empirical relationship that relates the effective viscosity of a suspension to the properties of the particles and the fluid.
When combining the mass ratio of stone powder and linseed oil as $1:1$,\footnote{We estimate here the order of Schmidt number, therefore the value of this ratio does not change our conclusion.} the effective viscosity is then,
\begin{equation}
     \mu_{\si{eff}}=\mu_{\si{f}} (1+2.5\frac{\rho_f}{\rho_f+\rho_s}),
\end{equation}
Substituting the given dynamic viscosity of linseed oil $\mu_{\si{f}}=0.055\, \si{Pa \cdot s}$ at the room temperature, that is $T=293.15\,\si{K}$, the density of linseed oil $\rho_f=0.93 \,\si{g/cm^3}$, the density of stone $\rho_s=2.5\,\si{g/cm^3}$, we get,
\begin{equation}
    \mu_{\si{eff}} = 1.68 \mu_{\si{f}}=9.24 \times 10^{-2} \, \si{Pa \cdot s},
\end{equation}
The effective kinematic viscosity is then estimated as, 
\begin{equation} 
\nu_{\si{eff}}= \frac{\mu_{\si{eff}}}{\rho_{\si{eff}}}\simeq 6.79 \times 10^{-5}\si{ m^2/s},\label{eq:effective_nu}
\end{equation}
where the effective fluid density is calculated as $\rho_{\si{eff}}\simeq 1360\,\si{kg/m^3}$.

It is important to note that the previously mentioned estimated Reynolds number is approximately $\mathcal{O}(10)$. Therefore, the Einstein equation condition may not hold. In this context, considering the order of the Reynolds number, a more precise effective kinematic viscosity does not alter our conclusion.

\added{Moreover, the diffusion coefficient of a spherical particle in a liquid can be estimated using the Stokes-Einstein equation,\citep{Einstein1905AP} which is written as, 
\begin{equation}
\kappa_{\si{eff}} = \frac{k_{Bol}  T } {6 \pi  \mu_{\si{f}}  r}.
\end{equation}
Here, $k_{Bol} = 1.38 \times 10^{-23}\si{ m^2 kg s^{-2} K^{-1}}$ is the Boltzmann constant; $T$ is the absolute temperature; $\mu_{\si{f}}$ is the dynamic viscosity of the liquid; and $r$ is the radius of the spherical particle.
We estimate here an order of the Schmidt number; therefore, we do not consider a nonspherical particle or a mixture of particle sizes where more complex models may be required.
Taking into account an average particle radius of $r=10 \,\si{\mu m}$ and a dynamic viscosity of the linseed oil
at room temperature, that is, $\mu_{\si{f}}=0.055\, \si{Pa \cdot s}$, the
mass diffusivity of the stone powder in the linseed oil can be estimated to be around $\kappa_{\si{eff}}\simeq 3.90\times 10^{-16}\si{ m^2/s}$. Finally, we have an estimation of Schmidt number as,
\begin{equation}
\si{Sc}=\frac{\nu_{\si{eff}}}{\kappa_{\si{eff}}}\simeq 1.74\times 10^{11} =\mathcal{O}(10^{11})
\end{equation}
This value is above the value of the low bound estimated from Fourier power spectrum. 
It is important to note that the above estimation assumes that the particles are small enough so that they do not interact with each other, which may not be the case for more concentrated suspensions or for particles with complex shapes.}

\clearpage

\section{Examination of Additional Images}\label{sec:SI_Images}

In this section, we examine two additional images, respectively, the painting \textit{Chain Pier, Brighton} by John Constable in 1826 and Jupiter Great Red Spot by Voyage 1 on 5 March 1979. The same analysis as for \textbf{\textit{The Starry Night}} is performed. The Kolmogorov-like $-5/3$ power spectra are evident since the turbulence-like pattern is well maintained in these two images.

\subsection{\textit{Chain Pier, Brighton} by John Constable}

\begin{figure}[!htb]
\centering
 \includegraphics[width=0.9\linewidth]{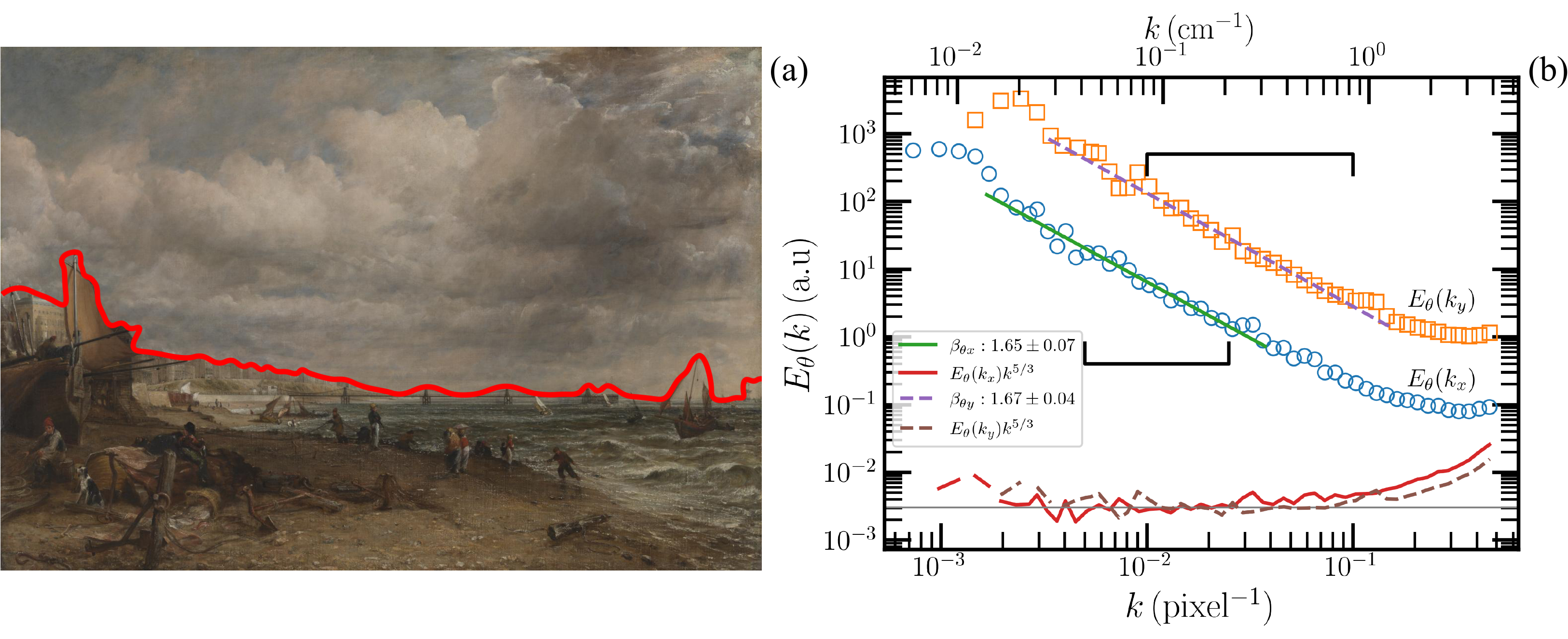}
 \caption{(a) \textit{Chain Pier, Brighton} painted by John Constable in 1827, obtained from {\href{https://www.tate.org.uk/art/artworks/constable-chain-pier-brighton-n05957}{https://www.tate.org.uk}}. The land 
 and the cloud sky is separated by the red line. (b) Experimental Fourier power spectrum $E_{\theta}(k)$ of \textit{Chain Pier, Brighton}. The green and purple dashed lines indicate power-law behaviors in the range $5\times 10^{-3}\,\si{pixel^{-1}}\lesssim k \lesssim 2.5\times 10^{-2}\,\si{pixel^{-1}}$ (i.e., $ 4.2\times 10^{-2}$\,\si{cm}$^{-1}\,\lesssim k\lesssim 2.1\times 10^{-1}\,\si{cm}^{-1}$) and $10^{-2}\,\si{pixel^{-1}}\lesssim k \lesssim 10^{-1}\,\si{pixel^{-1}}$ (i.e., $ 8.3\times 10^{-2}$\,\si{cm}$^{-1}\,\lesssim k\lesssim 8.3\times 10^{-1}\,\si{cm}^{-1}$) for the data in the horizontal and vertical directions, respectively. For display clarity, the curve of $E_{\theta}(k_y)$ has been shifted up vertically by multiplying a factor of 10. The red solid and brown dashed lines are compensated curves $E_{\theta}(k)k^{5/3}$ \added{to highlight the $-5/3$ scaling.}}
 \label{fig:chain}
\end{figure}\label{fig:SI_Chain_Pier_Brighton}

John Constable (11 June 1776 to 31 March 1837) was an English landscape artist associated with the Romantic tradition. He is primarily recognized for transforming the landscape painting genre. He conducted many observational studies of landscapes and clouds, aiming to be more scientific in capturing atmospheric conditions. The impact of his physical effects was often evident even in the large-scale paintings he displayed in London. The \textit{Chain Pier, Brighton} is one such painting, completed in 1826 and shown in 1827, in which the cloud/sky and beach/land are well separated. Unlike \textbf{\textit{The Starry Night}}, this painting lacks well-defined swirling patterns, but the clouds are rich of structures with different scales, resembling those frequently seen in the sky; see Fig.\,\ref{fig:SI_Chain_Pier_Brighton}\,(a).

A digital version of \textit{Chain Pier, Brighton} can be accessed from \href{https://www.tate.org.uk/art/artworks/constable-chain-pier-brighton-n05957}{https://www.tate.org.uk}. The dimensions of the image are $183\,\si{cm} \times 127\,\si{cm}$, equivalent to $1,536\,\si{pixel} \times 1,057\,\si{pixel}$, with a spatial resolution of approximately $0.12\,\si{cm/pixel}$. The original image is converted to gray-scale and treated as a scalar field. The Fourier power spectrum $E_{\theta}(k)$ for both horizontal ($x$) and vertical ($y$) directions is then calculated after excluding the land area, as shown in Fig.\,\ref{fig:SI_Chain_Pier_Brighton}\,(a). It is not surprising that the Kolmogorov-like $-5/3$ spectrum is evident in Fig.\,\ref{fig:SI_Chain_Pier_Brighton}\,(b) for both $E_{\theta}(k_x)$ and $E_{\theta}(k_y)$, as Constable accurately captured the cloud patterns.

\subsection{Jupiter Great Red Spot by Voyage 1}

\begin{figure}[!htb]
\centering
 \includegraphics[width=0.9\linewidth]{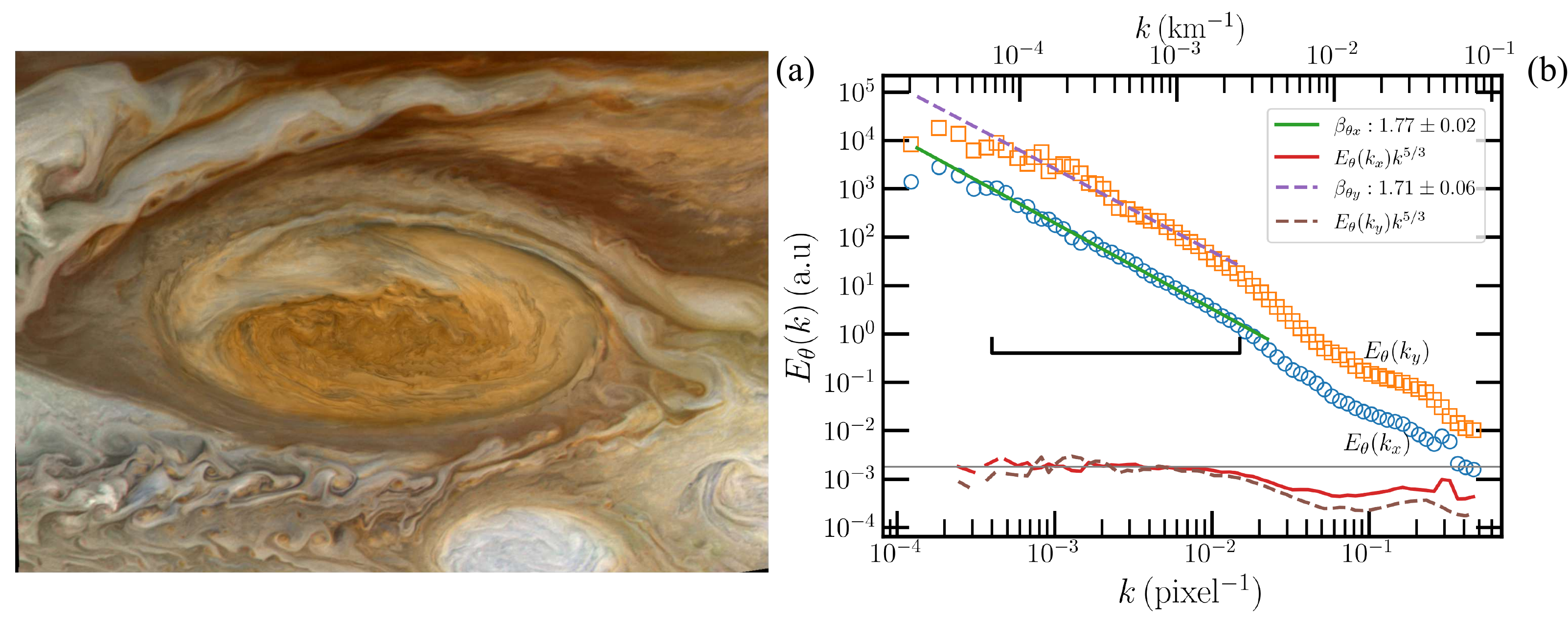}
 \caption{(a) 
 \textit{The Great Red Spot} obtained from {\href{https://www.planetary.org/space-images/voyager-1-view-of-the-great-red-spot}{https://www.planetary.org}} with a cropped size of $7,300\,\si{pixel} \times 5,050\,\si{pixel}$. \added{Courtesy of} NASA/JPL-Caltech/Bj\"{o}rn J\'{o}nsson. (b) Experimental Fourier power spectrum $E_{\theta}(k)$, in which the green and purple dashed lines indicate power-law behaviors in the range $4 \times 10^{-4}\,\si{pixel^{-1}}\lesssim k \lesssim 1.5\times 10^{-2}\,\si{pixel^{-1}}$ (i.e., $ 6.7\times 10^{-5}$\,\si{km}$^{-1}\,\lesssim k\lesssim 2.5\times 10^{-3}\,\si{km}^{-1}$) for the data in horizontal and vertical directions, respectively. For display clarity, the curve of $E_{\theta}(k_y)$ has been shifted vertically by multiplying a factor of 10. The red solid and brown dashed lines are compensated curves $E_{\theta}(k)k^{5/3}$ \added{to highlight the $-5/3$ scaling.}}
 \label{fig:Red_spot}
\end{figure}

The Great Red Spot is a long-lasting high-pressure area in Jupiter's atmosphere, creating the largest anticyclonic storm in the Solar System. It is the most distinctive feature on Jupiter, characterized by its red-orange hue. Situated $22$ degrees south of Jupiter's equator, it generates wind speeds up to $432\,\si{km/h}$. The Jupiter's Great Red Spot rotates counterclockwise with a period of approximately 4.5 Earth days with roughly $16,400\,\si{km}$ in width, making it $1.3$ times the diameter of Earth. The storm has persisted for centuries due to the absence of a solid planetary surface to create friction; gas eddies in the atmosphere continue for extended periods because there is no resistance to their angular momentum.\citep{Rogers2008JBAA}

A high-resolution image of the Great Red Spot can be found at \href{https://www.planetary.org/space-images/voyager-1-view-of-the-great-red-spot}{https://www.planetary.org}, with dimensions of $7,400\,\si{pixel} \times 5,550\,\si{pixel}$ and a spatial resolution of approximately $6\,\si{km/pixel}$. Captured by Voyager 1 on 5 March 1979, the image was taken using a green and violet filter mosaic with its narrow angle camera (NAC), covering the majority of the Great Red Spot. To highlight various details, the image's color, contrast, and sharpness have been enhanced. It is the highest resolution color data available for Jupiter before the Juno mission. A square region with a size of $7,300\,\si{pixel} \times 5,050\,\si{pixel}$ was cropped from the original by excluding the black edges of original stitched photo; see Fig.\,\ref{fig:Red_spot}\,(a). Visually, the Great Red Spot shows an ellipse-like pattern approximately with a major axis of $4,000\,\si{pixel}$ and a minor axis of $2,000\,\si{pixel}$, corresponding to $24,000\,\si{km}$ and $12,000\,\si{km}$. In addition to the Great Red Spot, very rich eddy-like structures can be seen, ranging in size from $50\,\si{pixel}$ to $2,000\, \si{pixel}$, corresponding to $300\,\si{km}$ to $12,000\,\si{km}$.

The raw image is converted to gray-scale and considered as a scalar field. The Fourier power spectrum $E_{\theta}(k)$ for both the horizontal $(x)$ and vertical $(y)$ directions is depicted in Fig.\,\ref{fig:Red_spot}\,(b). The Kolmogorov-like $-5/3$ spectrum is apparent in the range $4\times 10^{-4}\,\si{pixel^{-1}}\lesssim k \lesssim 1.5\times 10^{-2}\,\si{pixel^{-1}}$ (i.e., $ 6.7\times 10^{-5}$\,\si{km}$^{-1}\,\lesssim k\lesssim 2.5\times 10^{-3}\,\si{km}^{-1}$) in both horizontal and vertical directions. It is important to note that this scaling range aligns well with the measured spatial size of the eddy-like structures; see Fig.\,\ref{fig:Red_spot}\,(a). Similarly to our observations for \textit{\textbf{The Starry Night}}, both the spatial distribution and the relative intensity of these eddy-like structures adhere to the Richardson-Kolmogorov cascade picture. Here, the Kolmogorov-like $-5/3$ spectrum spontaneously emerged due to hydrodynamic interactions between different eddies.

\clearpage

\section{\textbf{\textit{The Starry Night}} in Everyday Life}


\begin{figure*}[!htb]
\centering
\includegraphics[width=0.7\linewidth]{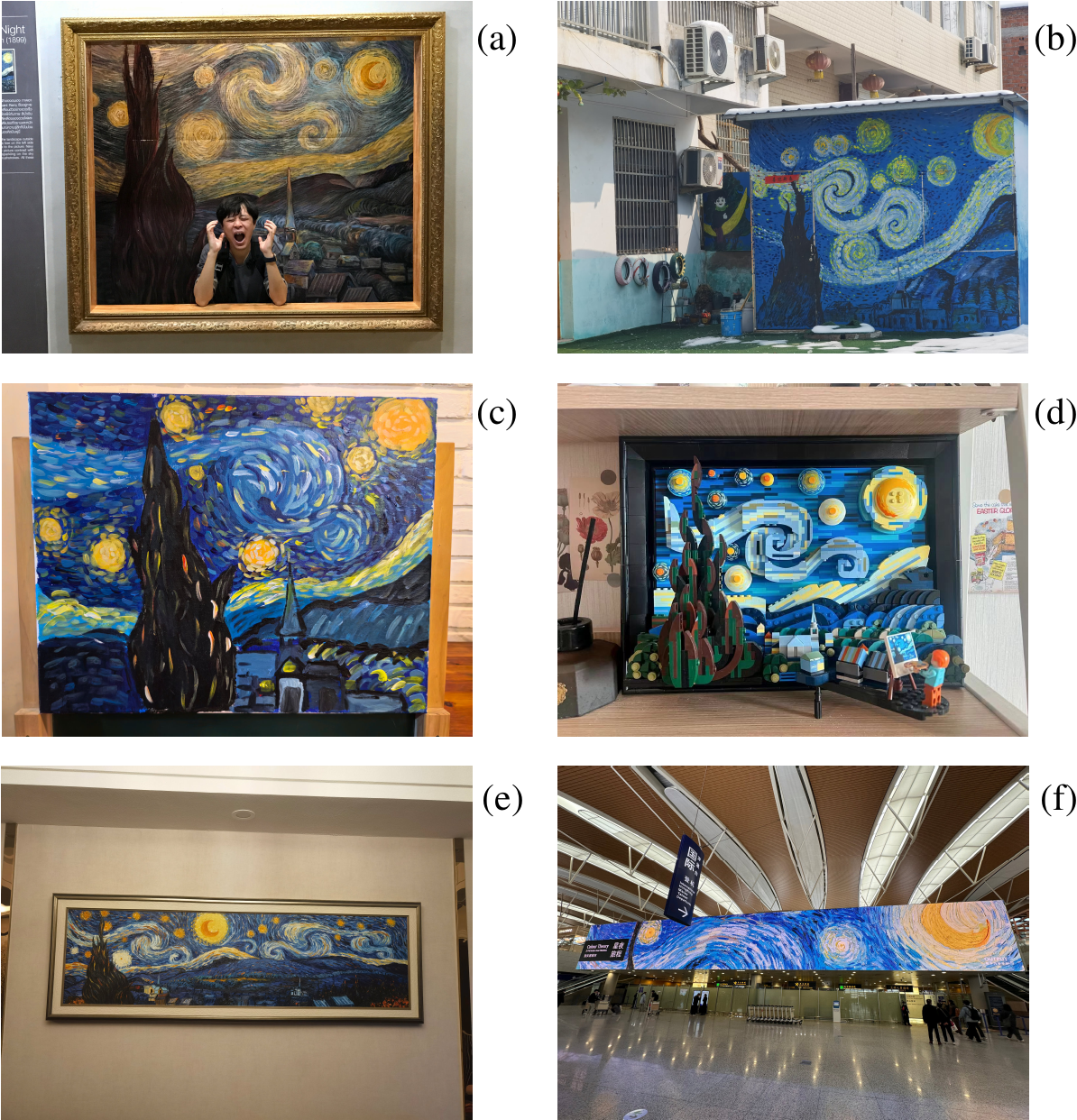}
 \caption{(Color online) Incorporating van Gogh's \textbf{\textit{The Starry Night}} into everyday life. (a) A man with a reproduction of \textbf{\textit{The Starry Night}} during an exhibition in Pattaya, Thailand. Photographed by X.L. on 19 February 2018. (b) A picture of \textbf{\textit{The Starry Night}} adorns the wall of a kindergarten in Randeng, a small town located in Fengyang County, Anhui Province, China. Photographed by Y.H. on 7 February 2024. (c) A practice painting of \textbf{\textit{The Starry Night}} by a 9-year-old girl, Ruoyi Xie on 2 August 2021. Photographed by X.L. in Fuzhou, Fujian Province, China, on 16 March 2024. (d) An image of a LEGO\textsuperscript{\textcopyright}
 jigsaw puzzle depicting \textbf{\textit{The Starry Night}}. Photographed by Ms. Xuan Lei in Shenzhen, Guandong Province, China, on 24 March 2024. (e) \textit{\textbf{The Starry Night}} graces a family home in Jiuquan, Gansu Province, China. Photographed by Ms. Xiangying Li, Jiuquan, Gansu Province, China, on 16 March 2024. (f) \textbf{\textit{The Starry Night}} on an advertisement board at the Pudong International Airport, Shanghai, China. Photographed by Mr. Fulian Gan on 7 April 2024. }
\label{fig:StarryNight_daily_Appendix}
\end{figure*}

\textit{\textbf{The Starry Night}} frequently appears in our everyday lives. Several instances are illustrated in Fig.\,\ref{fig:StarryNight_daily_Appendix}. For instance, Fig.\,\ref{fig:StarryNight_daily_Appendix}\,(a) shows an exhibition in Pattaya, Thailand, during Ms. X.L.'s visit on 19 February 2018. She captured this image with her husband standing in front of the replicated \textit{\textbf{The Starry Night}}. In Fig.\,\ref{fig:StarryNight_daily_Appendix}\,(b), a reproduced \textit{\textbf{The Starry Night}} decorates the wall of a kindergarten in Randeng, a small town in Fengyang County, Anhui Province, China, during Y.H.'s attendance at his niece's wedding on 7 February 2024. \textit{\textbf{The Starry Night}} is also cherished by children. For example, Fig.\,\ref{fig:StarryNight_daily_Appendix}\,(c) showcases a practice piece by a 9-year-old girl, Ms. Ruoyi Xie, on 2 August 2021. Meanwhile, Ms. Xuan Lei used a LEGO\textsuperscript{\textcopyright} jigsaw puzzle version of \textit{\textbf{The Starry Night}} to embellish her room in Shenzhen, Guangdong Province, China; see Fig.\,\ref{fig:StarryNight_daily_Appendix}\,(d). Several thousand kilometers from Shenzhen, Ms. Xiangying Li also selected a reproduced \textit{\textbf{The Starry Night}} to beautify her family home in Jiuquan, Gansu Province, China; see Fig.\,\ref{fig:StarryNight_daily_Appendix}\,(e). It is fascinating to observe \textit{\textbf{The Starry Night}} on an advertising board at the Pudong International Airport and Hongqiao International Airport, Shanghai, China; see Fig.\,\ref{fig:StarryNight_daily_Appendix}\,(f).
This advertisement promotes the artist Mr. Jesse Woolston's exhibition in Shanghai, China since the Mid-Autumn Festival, 10 September 2022. Mr. Woolston created a series of stunning works inspired by \textit{\textbf{The Starry Night}} and physics, which can be found at
{\href{https://www.youtube.com/watch?v=noycF6xQlBY}{https://www.youtube.com}} and 
{\href{https://www.tiktok.com/@jessewoolston\textunderscore/video/6933767826008329477}{https://www.tiktok.com}}.

We believe more examples can be found worldwide. We hope that the work showcased here will inspire the younger generation to participate in fundamental research, as sparking curiosity through captivating artwork is a crucial approach for advancing scientific progress. \added{Finally, we would like to quote the words directly from Ref.\,\onlinecite{Sherman2023PP}:}
\begin{displayquote}
\added{"We argue that although art has no systematic conventions for conveying knowledge in the way science does, the arts often play an important epistemic role in the production and understanding of scientific knowledge. We argue for what we call weak scientific cognitivism, the view that the production and distribution of scientific knowledge can benefit from engagement with art."}
\end{displayquote}

\clearpage
%


\end{document}